\documentclass[12pt]{article}
\usepackage[margin=1.2in]{geometry}
\usepackage{amsmath}
\usepackage{amssymb}
\usepackage{relsize}
\usepackage{bigints}
\usepackage{natbib}
\usepackage{graphicx}
\usepackage{xcolor}
\usepackage{bm}
\usepackage[labelfont={bf,small}, textfont=small, skip=1pt, font={stretch=1.2}]{caption}
\usepackage{upgreek}
\usepackage[utf8]{inputenc}
\usepackage{nameref}
\usepackage[colorlinks=true, allcolors=blue]{hyperref}
\usepackage{cleveref}
\usepackage{xfrac}
\usepackage{longtable}
\usepackage{setspace}
\usepackage{todonotes}

\def\correspondingauthor{\footnote{To whom correspondence must be addressed: bhanjan@che.iith.ac.in}}

\usepackage{mathtools,tikz,caption}
\usepackage{tikz}
\usepackage{authblk}

\usepackage{mathtools}
\usepackage{tikz-cd}
\usepackage{bm}
\usepackage{epstopdf}
\AppendGraphicsExtensions{.tif}

\title{Reverse segregation in dense granular flow through narrow vertical channel}
\author[1]{{Bhanjan Debnath}
\correspondingauthor{}}
\affil[1]{Department of Chemical Engineering, Indian Institute of Technology Hyderabad, Kandi, Sangareddy, Telengana 502284, India}

\def\md{\mbox{d}}
\date{}

\begin{document}
\maketitle

\begin{abstract}
\noindent
Controlling  flow-induced segregation in a granular mixture is highly relevant to many industrial settings. To enhance mixing or promote segregation, the continuous gravity flow of a bidisperse granular mixture through a series of narrow vertical channels with exit slots is investigated. The bidisperse mixture is composed of two different sizes of particles, but of the same density. In dense flow, segregation occurs, leading to formation of bands. The bands of large particles appear at a distance away from the walls. This finding is in contrast to that in shear-driven segregation in a dense flow where large particles segregate towards the walls. Using a phenomenological model, it has been shown that rolling and bouncing induced segregation is the dominant mechanism. When cylindrical inserts are placed to modify flow patterns, that significantly influences segregation patterns. The symmetrical placement of a cylindrical insert close to the exit slot vanishes the bands and enhances mixing. However, with two inserts placed symmetrically and close to the exit slot, the degree of segregation in the reverse direction is greatly enhanced compared to that without insert. In the former, small particles accumulate in thin regions adjacent to the walls, and large particles comprise the bulk of the domain and the flowing stream. The heap formation above the insert in a narrow channel, when the insert is close to the exit, enhances mixing in one configuration, whereas it amplifies reverse segregation in the other.

\end{abstract}

\section{Introduction}

Granular mixing and segregation are two important unit operations frequently encountered in the powder and grain industries \citep{ottino2001fundamental,rao2008introduction,bridgwater2012mixing}. It has long been known that granular mixtures constituted by particles of different sizes, densities, or surface properties  segregate under conditions, such as external vibration or shear \citep{ahmad1973observation,gray1997pattern,shinbrot1998reverse,ottino2000mixing,johnson2012grain}. Extensive research has been conducted on segregation in various flow configurations, including free surface flow inside a rotating tumbler \citep{gupta1991axial,mccarthy1996mixing,gray2001granular,jain2005regimes}, flow over an inclined chute \citep{khakhar1999mixing,wiederseiner2011experimental,tripathi2013density}, bounded heap flow \citep{fan2012stratification,fan2017segregation},  flow inside cylindrical Couette geometries  \citep{hill2008isolating,golick2009mixing}, and flow through a silo \citep{artega1990flow,samadani1999size,ketterhagen2008modeling}. In flow-induced segregation,  the pressure gradient induced by the gravity and the gradient of the shear rate can be dominant to influence segregation \citep{khakhar1999chaotic,gray2005theory,fan2011theory, umbanhowar2019modeling,jing2022drag,sahu2023particle,duan2024general}. It has been shown that the interplay among an advective flow field, diffusion due to grain temperature and  percolation determines  mixing and segregation \citep{khakhar1997transverse,fan2014modelling}. Despite a deep understanding, less is explored  how segregation can be controlled  in a flow configuration under the same operating conditions, such as the effect of overburden and  the mass flow rate. To address this former issue, one possible mechanism  can be the use of flow-modifying inserts \citep{cliff2021granular,irvine2023mu}.

Since many decades, flow of grains past inserts or obstacles was investigated \citep{wieghardt1975experiments,tuzun1985gravity,tuzun1985gravityII,tardos1998forces,chehata2003dense,gravish2014force,agarwal2021surprising}. An important aspect in such flows is to design inserts based on the forces (drag and lift) they encounter, and how the flow dynamics influence these forces \citep{ding2011drag,guillard2014lift,debnath2017lift,dhiman2020origin}. However, there is a little progress in this area of granular flow  due to their complex rheology and the lack of a suitable constitutive law \citep{debnath2022comparison}. Numerous studies showed complex flow  patterns and regimes in simple flow configurations that existing theories do not capture \citep{krishnaraj2016dilation,bharathraj2019cessation,dsouza2021dilatancy,debnath2022different,debnath2023cross}. Furthermore, mixing and segregation are complex processes. The coupling between the flow dynamics and segregation is not straightforward even in simple flow configurations \citep{guillard2016scaling,jing2022drag,yennemadi2023drag,liu2023coupled}.  Because of these reasons, segregation in flow past an insert  is less investigated. 

In flow past an insert, a  dense stagnant zone and shock wave arise in the upstream and a grain-free wake zone appears in the downstream. The sizes of the stagnant zone and the wake zone depend on the upstream velocity, the shape of the insert, and distances of the free surface and the exit slot from the insert \citep{wassgren2003dilute,cui2013gravity,mathews2022numerical,tregaskis2022subcritical}. If the insert is  cylindrical, the size of the wake zone reduces, adopting a triangular shape as the upstream velocity decreases \citep{chehata2003dense,cui2013gravity,chen2022flow,tregaskis2022subcritical}. There is an interesting observation related to the stagnant zone. A recent work \citep{mathews2022numerical} showed that the stagnant zone takes the shape of a stable heap due to a continuous flow. It  dissolves in absence of the flow. The height of the heap is a function of the diameter of the cylinder and  strongly depends on the solids fraction. However, it shows weak dependence on the grain diameter and the upstream velocity. In a dense flow, the maximum height of the heap is approximately the diameter of the cylinder. 

It can be speculated that if the cylinder's diameter is comparable to the  flow dimensions in a bounded dense flow, such heap formation  can significantly  alter the surrounding flow patterns. Therefore, it raises a question: can the extent of segregation be controlled by  modifying the flow patterns in a bounded dense flow using cylindrical inserts? 

Here we investigate a continuous flow through a series of narrow vertical channels with exit slots. The cylindrical inserts of diameter comparable to the narrow dimension of the channel are placed inside the bed at different locations to modify the flow patterns. We adopt a simulation based approach using the discrete element method (DEM) to simulate the flow.   We observe that without insert, large particles segregate away from the wall regions, in contrast to the outcome in a purely shear-driven process \citep{fan2011phase}. In the latter, large particles segregate towards the walls in a dense flow. Our phenomenological model reveals that rolling and bouncing induced segregation due to the free surface at the top is the dominant mechanism. The model explains how this mechanism drives large particles away from the walls in a narrow channel, which is different from free surface segregation \citep{fan2017segregation}. By placing inserts, we demonstrate that the degree of segregation can be controlled in a narrow channel. We show that the degree of segregation of large particles towards the central region is further enhanced when two inserts are placed symmetrically close to the exit slot, whereas the symmetrical placement of an insert near the exit slot reduces the degree of segregation. We present the details of the configurations with inserts and the DEM simulation in section~\ref{sec:DEM}, and report the results in section~\ref{sec:res} with a discussion and future scopes in section~\ref{sec:dis}.

\section{Flow configurations and DEM simulation}
\label{sec:DEM}

\begin{figure}
\centering
\includegraphics[width=0.8\linewidth]{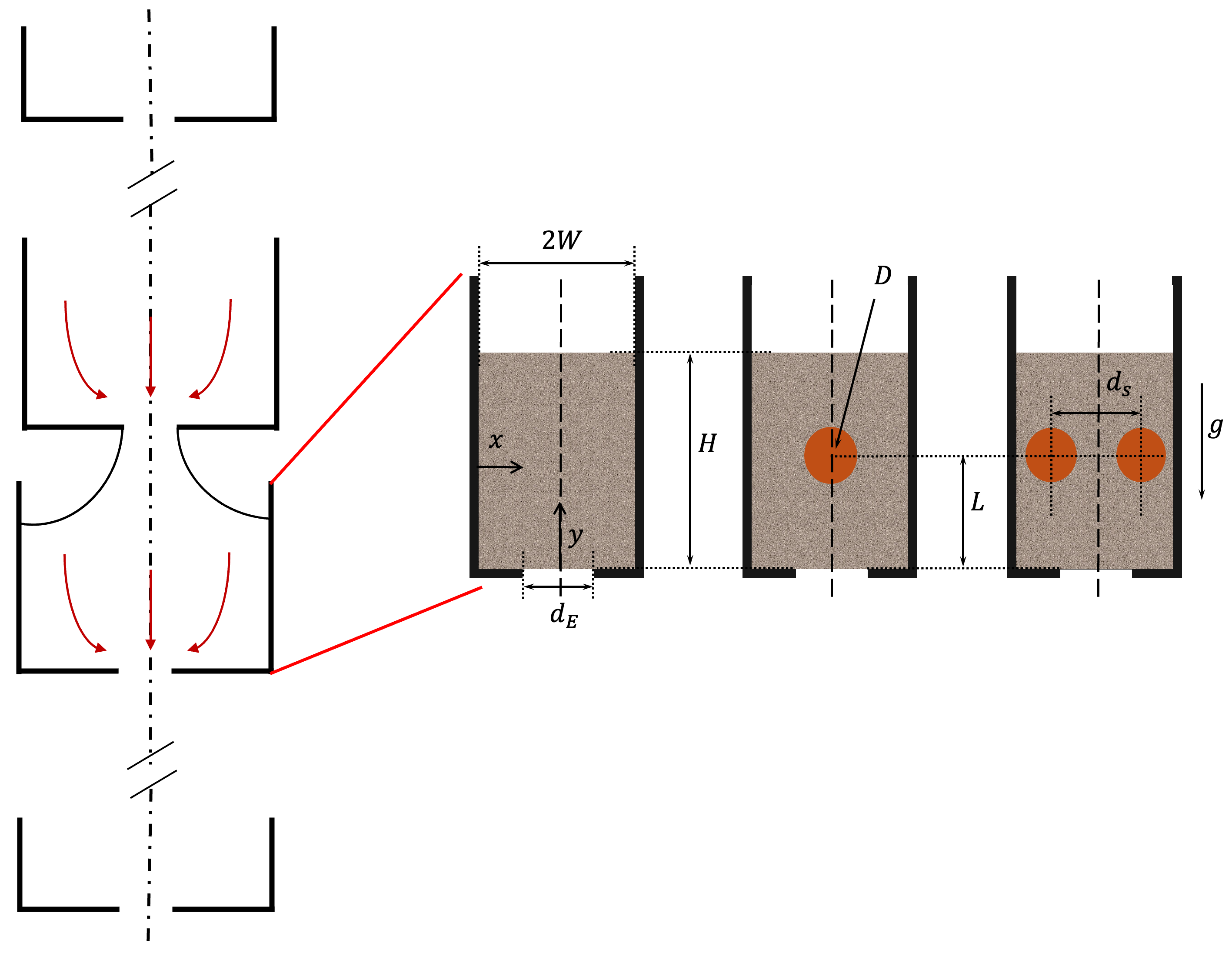}
\caption{The first panel is the schematic of gravity flow through a series of vertical channels. The other panels represent schematics of one channel without and with inserts (of diameter $D$) at different locations.}
 \label{fig:schematic}
\end{figure}

The discrete element method (DEM) is used to simulate the continuous gravity flow through a series of vertical channels (figure~\ref{fig:schematic}).
The dimensions of one channel are $2W \times H \times B$ in the $x-$, $y-$ and $z-$ directions, respectively. Instead of simulating flow through a series of channels, we employ periodic boundary conditions in the flow ($y-$) and vorticity $z-$ directions to avoid simulating large number of particles. The flow is bounded by two flat frictional walls at a distance $2W$. To control the flow in the $y-$ direction, there is a flat bottom with an exit slot of width $d_E$.

To prepare the granular column, the exit slot is closed and the simulation box is filled with particles by uniformly pouring them from the top. The particles are spherical and of two different diameter $d_p$ and $d_p/2$. Here $d_p$ is denoted as the  nominal particle diameter. Approx. $10^5$ particles  are poured with number fraction 1:1 for $d_p$ and $d_p/2$. As the particle  density is the same for both  large and small particles,  it implies that the mass ratio of  large to  small particles is 8:1. After the particles settle under gravity, the height of the static bed attained is $H^{\star} \sim 50\, d_p $ for  $2W = 50\,d_p$ and $B = 20 \, d_p$. To initiate the flow, the exit slot of width $d_E = 10\,d_p$ is opened, $H$ is set to $60\,d_p$,  and the periodic boundary conditions are imposed in the $y-$ and $z-$ directions. 

To investigate the role of inserts, two new configurations are generated where cylindrical inserts are placed at a height $L$ from the bottom of the channel:
\begin{enumerate}
\item one  insert with its axis parallel to the $z-$ direction is  placed symmetrically with respect to the central plane of the  channel (figure~\ref{fig:schematic}),

\item two  inserts with their axes parallel to the $z-$ direction  are  placed symmetrically with respect to the central plane of the channel (figure~\ref{fig:schematic}). The symmetrical placement of two inserts yields to a center-to-center distance $d_s = W$.
\end{enumerate}

In both configurations, the inserts are kept stationary.  The diameter of the insert is $D$ and its length is equal to the depth $B$. The outer surface of an insert is roughened by coating it with rigid spherical particles of diameter $d_p$ arranged in close-packed hexagonal lattice. To place an insert inside the channel, particles are removed from the cylindrical region of dimensions 0.5 $d_p$ greater than the dimensions of the insert to avoid overlap between  insert-particles and surrounding active particles. 

In the present work,  we focus on inserts influencing mixing and segregation in a vertical channel under the same operating conditions such as the overburden and the mass flow rate.  That's why, we do not vary the initial filled height $H^{\star}$, $2W$, $B$ and $d_E$. The only parameters left are $D$ and $L$. At present, we fix the value of $D = 15\,d_p$  and report the results for three values of $L = $ 15 $d_p$, 30 $d_p$, and 45 $d_p$. 

The linear spring-dashpot contact force model is used to simulate particle--particle, particle--wall, and particle--insert interactions. In all interactions, the model parameters such as the spring constants ($k_n$, $k_t$), damping constants ($\xi_n$, $\xi_t$), and coefficient of friction ($\mu_p$) are set the same. The details of the DEM and values of the model parameters are described in our earlier work \citep{debnath2022comparison,
debnath2022different}. The length is scaled by $d_p$, time by $\sqrt{d_p/g}$, velocity by $\sqrt{g \, d_p}$, force by $\rho_p \, g \, d_p^3$, and stresses by $\rho_p \, g \, d_p$. Without loss of generality, $d_p$, $\rho_p$ and $g$ can be set to 1. To perform spatial  averaging of the  flow properties locally, the simulation box is divided into bins of dimensions 2 $d_p \times$ 2 $d_p$ in the $x-$ and $y-$ directions which span the depth $B$ in the $z-$ direction.

\section{Results}
\label{sec:res}

In the present work, the number fraction will provide an estimate of the extent of segregation. For example, the number fraction $n_f^l$ of large particles is defined as the ratio of the number of large particles to the total number of particles in an averaging bin. Note that the flow configuration is a flat-bottomed bin with an exit slot. To generate a continuous flow, the simulation box dimension in the flow $y-$ direction is  chosen higher than the height of the static bed.   Therefore, the formation of a free surface at the top and stagnant zones at the corners of the bottom cannot be avoided in the flow configurations investigated here. The average solids fraction in the flowing state is $\bar{\phi} \sim 0.59$.   

\subsection{With no insert}
\label{sec:res_1}

\begin{figure}
\centering
\includegraphics[width=0.8\linewidth]{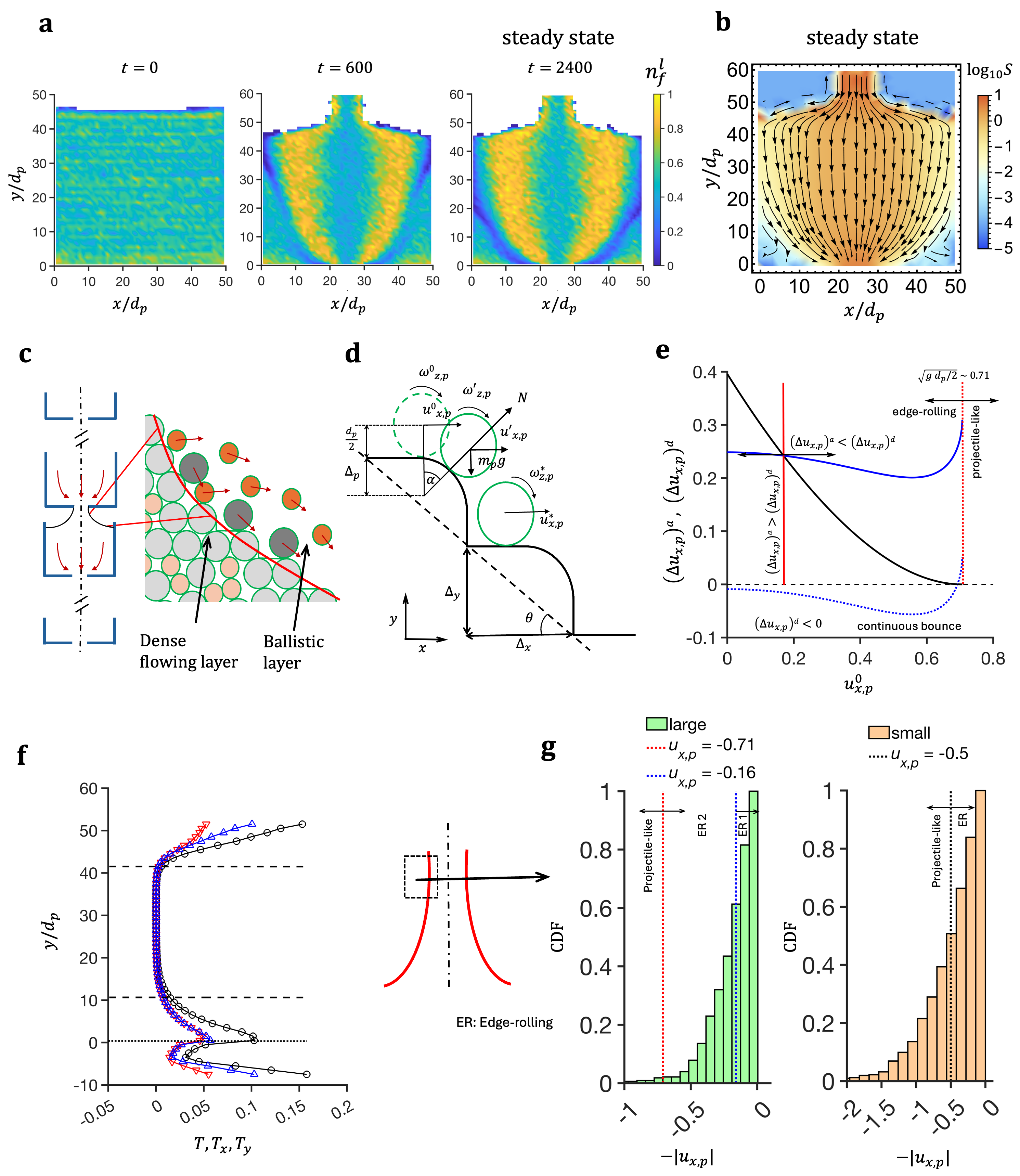}
\caption{With no insert: (a) The spatial distribution of large particles at different time points. The background color represents the number fraction $n_f^l$ of large particles. (b) Streamlines of the flow at steady state. The background color represents the value ${\rm{log}_{10}}S$, where $S = (u_x^2 + u_y^2)^{1/2}$ is the magnitude of the flow. (c) Schematic of particles rolling and bouncing over the free surface. (d) Schematic of the mechanism of single particle rolling over stairs. (e) The increment $(\Delta u_{x,p})^a$ and decrease $(\Delta u_{x,p})^d$ in the horizontal velocity as functions of $u_{x,p}^0$ due to the acceleration and deceleration processes, respectively. The black curve represents $(\Delta u_{x,p})^a$, and the blue curves represent  $(\Delta u_{x,p})^d$. The blue solid curve is for $\theta = 12^{\text o}$ and the blue dotted curve for $\theta = 30^{\text o}$. (f) Variation of the granular temperature with the vertical direction $y-$ at the mid-plane of the channel. The black, red, and blue curves represent $T$, $T_x$, and $T_y$, respectively. (g) The cumulative density function (CDF) of the horizontal velocity of individual particles (small and large) leaving from the left edge of the feed stream.
}
 \label{fig:noinsert}
\end{figure}

Figure~\ref{fig:noinsert}a shows the number fraction distribution of large particles ($n^l_f$) at different times. Initially, the mixture is almost uniform. The average solids fraction, which is the ratio of the volume of all particles to the volume of the region in the channel that the material occupies, is  significantly higher ($\bar{\phi} \sim 0.59$), indicating a dense flow.  As time progresses, segregation occurs and  bands of large particles appear in  regions offset to the central zone and the walls. The central zone and the stagnant zones remain less affected by segregation. The material in the central zone moves as a plug except close the free surface and the exit slot, and the velocities are negligible in the stagnant zones (figure~\ref{fig:noinsert}b). At a steady state, a striking observation is that small particles accumulate near the walls. There are concentrated layers of small particles moving slowly close to the walls and just above the stagnant zones  (figures~\ref{fig:noinsert}a,b). It appears that the small particles are forming an interface between the stagnant zone and the flowing zone close to the exit slot. These observations are opposite to that in shear-driven segregation. 

Size segregation in a vertical channel has been shown to be a shear-driven process \cite{fan2011phase}. In the former, large particles accumulate near the walls and small particles between the central zone and the walls in dense flows when $\bar{\phi}$ is high. In contrast, this trend reverses when the material is loose under dilute flow conditions, with large particles segregating towards the center of the channel \cite{fan2011phase}. In another context, if there is an inclined free surface at the top (see figure~\ref{fig:noinsert}b), it can increase the probability of accumulation of large particles near the walls \citep{samadani1999size, fan2012stratification}. However, the bands of large particles appear at a distance away from the walls in the present flow configuration (figure~\ref{fig:noinsert}a). This raises the question: which effect  dominates the shear and the free surface flow, leading to the opposite trend in a dense flow configuration?  

We hypothesize that rolling and bouncing motion of particles above the free surface can affect segregation trends in the following manner.
The particles escaping the inlet stream hit the free surface and bounce. Small particles are more likely to bounce more and can reach the walls in a narrow channel, whereas large particles tend to roll over the free surface. If large particles stop in between while rolling before reaching the walls and advect downward with the flowing stream, this mechanism can result in the segregation of large particles away from the walls. Using a phenomenological model, we explain how bouncing and rolling of particles over an inclined free surface can induce reverse segregation.

We assume that there are two flowing layers close to the free surface (see figure~\ref{fig:noinsert}c): (i)  dilute ballistic layer above the free surface where particles are not at sustained contacts, and (ii) relatively dense flowing layer of particles at sustained contacts just below the free surface. In the ballistic layer, particles can bounce or roll over the free surface.  We imagine the free surface as a series of stairs with rounded edges, continuously forming and vanishing with the flow. The average height and width of the stairs are equivalent to the size of the particles. Assuming the ballistic layer to be very dilute, we can consider the motion of individual particles above these stairs.

\subsubsection{Rolling and bouncing motion of a particle over stairs with rounded edges}
\label{sec:res_1_1}

For the sake of simplicity, the two dimensional motion of one particle of diameter $d_p$ and mass $m_p$ above the stairs with rounded edges is considered (see figure~\ref{fig:noinsert}d). We adapted the model of \cite{yan2007dynamics}, which is for a viscous ball rolling down stairs, and modified it for our system. The particle moves forward due to its initial horizontal velocity. We assume pure rolling and neglect sliding. The particle  gains acceleration when it rolls over a rounded edge. The particle can bounce for a while when it hits the next horizontal surface and dissipate its vertical velocity. This effect has not been considered in the model of \cite{yan2007dynamics}. The particle then experiences deceleration while rolling over the horizontal surface before reaching the next edge. The competition between these two contributions  determines whether the particle  moves with a constant velocity or stop after traversing a distance.

We will derive the increment in the velocity due to the acceleration process when a particle rolls over an edge (see figure~\ref{fig:noinsert}d). Consider the particle starts rolling over a rounded edge with horizontal velocity $u_{x,p}^0$ and angular velocity $\omega_{z,p}^0$. When the particle reaches an angle $\alpha$ being at contact with the edge, $u^{\prime}_{p}$ and $\omega^{\prime}_{z,p}$ are its translational and angular velocities. The energy balance of the particle reads as
\begin{equation}
    \frac{1}{2}\,m_p \, [u^{0}_{x,p}]^2 + \frac{1}{2} \, I_p \, [\omega^{0}_{z,p}]^2  = \frac{1}{2}\,m_p \, [u^{\prime}_{p}]^2 + \frac{1}{2} \, I_p \, [\omega^{\prime}_{z,p}]^2 - m_p \, g \, \Big(\frac{d_p}{2} +  \Delta_p \Big) \, (1 - {\rm{cos}} \, \alpha) + D_r. 
    \label{eq:preacc1}
\end{equation}
Here $g$ is the gravitational acceleration, $\Delta_p$ represents a distance (see figure~\ref{fig:noinsert}d), and $D_r$ is the energy dissipated due to rolling friction \citep{tabor1955mechanism}. We will discuss the contribution of $D_r$ in a short while. Using (\ref{eq:preacc1}), the assumption of pure rolling, and $I_p = (2/5) \, m_p \, (d_p/2)^2$, we can show that the particle velocity   $u^{\prime}_p$ in the limit $\Delta_p \ll d_p/2$ is 
\begin{equation}
    [u^{\prime}_p]^2 = [u^0_{x,p}]^2 + \frac{10}{7} \, g \, \frac{d_p}{2} \, (1 - {\rm{cos}} \, \alpha) - \frac{10}{7} \, \frac{D_r}{m_p}.
    \label{eq:acc1}
\end{equation}
When the particle is at the edge, the force balance in the normal and tangential directions, and the torque balance at the point of contact are, respectively,
\begin{equation}
  N - m_p \, g \, {\rm cos} \, \alpha + m_p \, \frac{[u^{\prime}_p]^2}{d_p/2}  = 0 \, ; \; \; m_p \, \frac{\md u_p^\prime}{\md t} = - T_r \, ; \; \; I_p \, \frac{\md \omega_{z,p}^\prime}{\md t} = T_r \, \frac{d_p}{2} - M_r,
    \label{eq:preacc1_1}
\end{equation}
where $T_r$  is the tangential force opposite to the direction of motion, and $M_r = \mu_r \, u_p^\prime \, N$ \citep{brilliantov1998rolling} is the rolling friction torque, where $\mu_r$ is the coefficient of rolling friction. It is reasonable to assume that the dissipation $D_r \leq T_r \, (\pi\, d_p/4)$, where the particle over the circular edge can travel a maximum distance equal to one-quarter of the perimeter of the circular edge before it loses contact with the edge. The diameter of the circular edge is $d_p$, as the rounded edge consists of particles with diameter $d_p$.  For the case of pure rolling and using (\ref{eq:preacc1_1}), it can be shown that  
\begin{equation}
    \frac{10}{7} \, \frac{D_r}{m_p} \leq \frac{50}{343} \, \pi \, d_p \, \mu_r \, u_p^\prime \, \Bigg[ g \, {\rm{cos}} \, \alpha -  \frac{[u_p^\prime]^2}{d_p/2} \Bigg]  
    \label{eq:preacc1_4},
\end{equation}
\cite{brilliantov1998rolling} showed that the value of $\mu_r = 10^{-6}$ s is very small for a steel ball rolling on a flat non-corrugated surface. Treating the rounded edge in the current problem as equivalent to a flat surface, the third term on the right-hand side of (\ref{eq:acc1}) appears to be negligible due to (\ref{eq:preacc1_4}). Therefore, in the subsequent steps of the analysis, we neglect the term $(10/7)\,(D_r/m_p)$ in (\ref{eq:acc1}).     

When the particle leaves the edge at an angle $\alpha = \alpha_c$, the normal force $N = 0$. Using (\ref{eq:acc1}) (neglecting $(10/7)\,(D_r/m_p)$) and (\ref{eq:preacc1_1}), the value of $(\rm{cos} \, \alpha_c)$ and the particle velocity  $[u^{\prime}_p]_c$ after loosing contact with the  edge  can be obtained as
\begin{align}
    {\rm cos} \, \alpha_c &= \Bigg(\frac{7\, [u^0_{x,p}]^2}{17 \, \frac{d_p}{2} \, g} + \frac{10}{17} \Bigg), \label{eq:acc21} \\
    [u^{\prime}_p]_c^2 &= \frac{1}{17} \Big(7 \, [u^0_{x,p}]^2 + 10 \, \frac{d_p}{2} \, g \Big).
    \label{eq:acc22}
\end{align}
Here (\ref{eq:acc21}) implies that  $u^0_{x,p} < (g\,d_p/2)^{1/2}$. If $u^0_{x,p} \geq (g\,d_p/2)^{1/2}$, the particle does not roll over the edge. It leaves the edge as a projectile, bounces and rolls on horizontal surfaces of the stairs. After cycles of energy dissipation through bouncing and rolling, the horizontal velocity eventually decreases to $ < (g\,d_p/2)^{1/2}$. Setting $d_p = 1$ and $g = 1$, the maximum value of  $u^0_{x,p}$ below which the particle rolls over the edges is  $(g\,d_p/2)^{1/2} \sim 0.71$.

Therefore, $u^0_{x,p} = (g\,d_p/2)^{1/2} = 0.71$ marks the boundary between two regimes of edge-rolling and projectile-like motions. Now, we consider $u^0_{x,p} < (g\,d_p/2)^{1/2}$, and analyze the scenario of the particle landing on the next stair after rolling over an edge. 

When the particle lands on the horizontal surface, it can bounce due to its non-zero vertical velocity for a short period of time. Upon bouncing, it would not be in continuous contact with the horizontal surface. The particle dissipates its vertical velocity after a few subsequent bounces due to inelastic collisions. We assume that the collisions during bouncing are not oblique, which may not be valid in the current scenario. However, to avoid complexities in the formulation, we neglect the effects of oblique impacts, for example, dissipation of the horizontal and angular velocities. 

The vertical velocity reduces to $u^\star_{y,p}|_n = (e_p)^n \, [u^\prime_{y,p}]_c$ after subsequent bounces $n$, where $e_p $ is the coefficient of restitution. The horizontal distance $\ell_b$ that the particle traverses while dissipating its vertical velocity is
\begin{equation}
 \ell_b = \sum_{n = 0}^{\infty} \; [u^\prime_{x,p}]_c \, \frac{2 \, (e_p)^n \,  [u^\prime_{y,p}]_c}{g},
    \label{eq:dis_n_bounce1}
\end{equation}
where $[u^\prime_{x,p}]_c = [u^\prime_p]_c \; {\rm cos} \, \alpha_c$ and $[u^\prime_{y,p}]_c = [u^\prime_p]_c \; {\rm sin} \, \alpha_c$. As $e_p < 1$, (\ref{eq:dis_n_bounce1}) is a geometric series which reduces to 
\begin{equation}
 \ell_b = \frac{([u^\prime_{p}]_c)^2 \, {\rm sin} \, (2 \alpha_c)}{g \, (1 - e_p)}.
    \label{eq:dis_n_bounce2}
\end{equation}
The distance $\ell_b \rightarrow 0$ for  $\alpha_c \rightarrow 0$, irrespective of $e_p$.  

When the vertical velocity dissipates completely, the particle starts rolling over the horizontal surface with horizontal and angular velocities $u_{x,p}^\star$ and $\omega_{z,p}^\star$, respectively.
When the vertical velocity is negligible, the energy balance can be written as
\begin{equation}
  \frac{1}{2}\,m_p \, [u^{\prime}_{x,p}]_c^2 + \frac{1}{2} \, I_p \, [\omega^{\prime}_{z,p}]_c^2  \approx  \frac{1}{2}\,m_p \, [u^{\star}_{x,p}]^2 + \frac{1}{2} \, I_p \, [\omega^{\star}_{z,p}]^2.
    \label{eq:eb_land1}
\end{equation}
Neglecting sliding friction, assuming pure rolling and using (\ref{eq:eb_land1}), it can be shown that 
\begin{equation}
    [u_{x,p}^\star]^2 = \frac{5}{7} \, [u^\prime_{x,p}]^2_c + \frac{2}{7} \, [u^\prime_p]^2_c,
    \label{eq:acc3}
\end{equation}
where $[u^\prime_{x,p}]_c = [u^\prime_p]_c \, \,  {\rm{cos} \, \alpha_c}$. Therefore, we can define an increment function for the horizontal velocity as (and using (\ref{eq:acc21}) and (\ref{eq:acc22}))
\begin{align}
      (\Delta u_{x,p})^a  = u_{x,p}^\star  - u^0_{x,p}   = \Bigg( \frac{7 [u^0_{x,p}]^2 + 10 \, \frac{d_p}{2} \, g}{17}\Bigg)^{1/2} \, \, \Bigg( \frac{5 \, \rm{cos}^2\, \alpha_c + 2}{7} \Bigg)^{1/2} \, - \; u^0_{x,p}.
      \label{eq:acc_incre}
\end{align}
Note that  (\ref{eq:acc_incre}) is  a function of $u^0_{x,p}$. The increment $(\Delta u_{x,p})^a$  is plotted in figure~\ref{fig:noinsert}e as a function of $u^0_{x,p}$.  

Now, we will discuss the decrease in the horizontal velocity. As mentioned earlier, $u_{x,p}^\star$ must decrease due to the rolling friction \citep{tabor1955mechanism} as the particle rolls over the horizontal surface towards the next edge (see figure~\ref{fig:noinsert}d).  Using the rolling friction torque $M = (\mu_{r} \, u^\star_{x,p} \, m_p \, g)$ \citep{brilliantov1998rolling}, assuming pure rolling, neglecting sliding friction, and using the balances of torques and forces, it can be shown that 
\begin{equation}
    [u^\star_{x,p}]^d (t) = u^\star_{x,p} \, {\rm exp} \, \Big( - \frac{5}{7} \, \frac{\mu_r \, g}{d_p/2} \, t \Big).
    \label{eq:dec1}
\end{equation}
As (\ref{eq:dec1}) is an exponential function, we can define the decrease  in the horizontal velocity due to the deceleration process as
\begin{equation}
      (\Delta u_{x,p})^d =  \Big( \frac{5}{7} \, \frac{\mu_r \, g}{d_p/2} \Big) \, \ell_x,
      \label{eq:dec_decre}
\end{equation}
where $\ell_x = f(\Delta_y, \Delta_x, \ell_b, u^0_{x,p})$ is the distance that the particle rolls over the horizontal surface, and $\Delta_y$ and $\Delta_x$ are the heights and breadths of the stairs, respectively.  

The value of $\Delta_y$ is crucial. If $\Delta_y \leq (d_p/2) \, (1 - {\rm cos} \, \alpha_c)$, the particle touches the horizontal surface before leaving the edge,  and the dynamics would be completely different. It is difficult to comment on the minimum value of $\Delta_y$, as it would depend on the  local arrangement of stair-particles. We assume a hexagonal arrangement and set  $\Delta_y = \sqrt{3} d_p/2$, which is higher than $d_p/2$. Therefore, the issue of particle touching the horizontal surface before leaving the edge would not be relevant here. The value of $\Delta_x = \Delta_y/{\rm tan} \, \theta$, where $\theta$ is  the angle of inclination of the free surface. Therefore, the value of $\ell_x$ would be
\begin{equation}
    \ell_x = \Delta_x  - \Bigg[ \frac{d_p}{2} \, {\rm sin} \, \alpha_c + \frac{[u^\prime_{x,p}]_c \, \Big(\sqrt{[u^\prime_{y,p}]^2_c  + 2\, g\,  \ell_y} - [u^\prime_{y,p}]_c \Big) }{g} 
 \Bigg] - \ell_b ,
    \label{eq:dec_len}
\end{equation}
where $\ell_y = \Delta_y - (d_p/2) \, (1 - {\rm cos} \, \alpha_c)$,  $[u^\prime_{x,p}]_c = [u^\prime_{p}]_c \, {\rm cos} \, \alpha_c$ and $[u^\prime_{y,p}]_c = [u^\prime_{p}]_c \, {\rm sin} \, \alpha_c$. On the right-hand side of (\ref{eq:dec_len}), the second term inside the square brackets corresponds to the horizontal distance the particle travels before landing, and the description of the third term $\ell_b$ in (\ref{eq:dec_len}) has been provided earlier (see (\ref{eq:dis_n_bounce2})). The former mechanism would be valid only for $\ell_x > 0$. If $\ell_x \leq 0$, then $(\Delta u_{x,p})^d < 0$, which is unacceptable, implying a continuous bouncing motion of the particle.  

The function $(\Delta u_{x,p})^d$ in  (\ref{eq:dec_decre}) is plotted in figure~\ref{fig:noinsert}e  (blue curves) for two different values of $\theta = 12^{\text o}$ and $30^{\text o}$, setting $\Delta_y = \sqrt{3} d_p/2$, $\mu_r = 0.07$ s and $e_p = 0.7$. The choice of $\Delta_y$ has been discussed earlier. A large value of $\mu_r$ is expected for a spherical particle interacting with a bumpy-frictional surface or a nearly spherical/irregular-shaped particle interacting with a flat-frictional surface \citep{wensrich2012rolling, cross2015effects,zhang2018size,tripathi2021quantitative}.  For $\theta = 12^{\text o}$, $(\Delta u_{x,p})^d > 0$ and $(\Delta u_{x,p})^d$ intersects $(\Delta u_{x,p})^a$. 
For a higher angle of inclination $\theta = 30^{\text o}$ and $(\Delta u_{x,p})^d < 0$, a particle bounces continuously. Therefore, there are four regimes (figure~\ref{fig:noinsert}e):

\begin{enumerate}
    \item Edge-rolling regime 1: In this regime, $u^0_{x,p} < (g\,d_p/2)^{1/2} \sim 0.71$ and $(\Delta u_{x,p})^d \leq (\Delta u_{x,p})^a$. The left of the intersection point in figure~\ref{fig:noinsert}e implies continuous rolling motion of the particle over the stairs. 
    \item Edge-rolling regime 2: In this regime, $u^0_{x,p} < (g\,d_p/2)^{1/2} \sim 0.71$ and $(\Delta u_{x,p})^d > (\Delta u_{x,p})^a$. The right of the intersection point in figure~\ref{fig:noinsert}e implies that the particle rolls over the stairs, but it stops after traversing a distance.
    \item Projectile-like regime: In this regime,  $u^0_{x,p} \geq (g\,d_p/2)^{1/2} \sim 0.71$. The particle leaves  stair-edges as a projectile and after subsequent bounces, its motion falls under the edge-rolling regimes.
    \item Continuous bounce: In this regime ($\ell_x \leq 0$), the particle bounces continuously and never falls under the edge-rolling regimes.  
\end{enumerate}

These regimes explain the reverse segregation pattern shown in figure~\ref{fig:noinsert}a  as follows.

\subsubsection{Model explains how bouncing and rolling induce reverse segregation}
\label{sec:res_1_2}

The central zone remains less effected by segregation, and the granular temperature is negligible in this zone  (between two black dashed horizontal lines in figure~\ref{fig:noinsert}f). The granular temperature is a measure of the velocity fluctuations, defined as $T = T_x + T_y + T_z$, where $T_x = (u_x^\prime)^2/3$, $T_y = (u_y^\prime)^2/3$, and $T_z =  (u_z^\prime)^2/3$ are the contributions of the velocity fluctuations in the $x-$, $y-$, and $z-$ directions, respectively \citep{debnath2024dense}. However, near the exit location and the feed stream, $T$ is not negligible (figure~\ref{fig:noinsert}f). (Note that the zone in the negative $y$ in figure~\ref{fig:noinsert}f represents the part of the feed stream due to the periodic boundary in the $y-$ direction). Due to the converging flow near the exit slot, $T_x$ is comparable to $T_y$ (figures~\ref{fig:noinsert}b,f). The former implies that the horizontal velocity of the individual particles is not negligible in the exit location and the feed stream. As $T$ is high at the feed stream location and the feed stream has traction-free edges, particles can escape from these edges. Those particles leaving the edges either roll down or bounce over the free surface. Then, this scenario is equivalent to what has been discussed in the previous sub-section \ref{sec:res_1_1}.

During the onset of segregation between time points $t = 3$ and $t = 600$, the horizontal velocity distributions of individual particles (small and large) that escape from the left edge of the feed stream are shown in figure~\ref{fig:noinsert}g. (Because the flow is symmetric, only the left side is shown.) A negligible fraction of large particles of diameter $d_p = 1$ has velocity $ < -(g\, d_p/2)^{1/2} = - 0.71$ (figure~\ref{fig:noinsert}g).  Hence, large particles leaving the traction-free surface move by rolling along the free surface and fall under edge-rolling regimes.  As the diameter of small particles is $d_p/2 = 1/2$, the critical value above which they escape as projectiles is $(g\, d_p/4)^{1/2} = 0.5$ (see sub-section~\ref{sec:res_1_1}). Figure~\ref{fig:noinsert}g shows that about $\sim$50\% of small particles exceed this critical value and, therefore, escape as projectiles. It indicates that small particles are more likely to reach the walls via bouncing motion in a narrow channel. But bouncing alone cannot explain why only small particles accumulate near the walls. Large particles can also reach the walls by rolling over the free surface, which is less likely in the current flow configuration for the following reason.  

As mentioned in sub-section~\ref{sec:res_1_1}, there are two regimes when a particle rolls. In the edge-rolling regime 1, $(\Delta u_{x,p})^d \leq (\Delta u_{x,p})^a$ and a particle rolls down continuously. In the edge-rolling regime 2, $(\Delta u_{x,p})^d > (\Delta u_{x,p})^a$ and a particle stops after traversing a distance. In the current flow configuration, it is observed that $\theta \sim 12^{\text o}$, for which the boundary between edge-rolling regimes 1 and 2 is marked at $u^0_{x,p} \sim 0.16$ when $d_p = 1$. The value of $\Delta_x \sim 4\,d_p$ for $\Delta_y = \sqrt{3}d_p/2$ and $\theta = 12^{\text o}$. This value of $\Delta_x$ is approximately one-sixth of the half-width $W$ of the channel. For this small value of $\theta$, figures~\ref{fig:noinsert}e,g show that the motion of about $\sim$60\% of large particles falls under the edge-rolling regime 2. Therefore, it is more likely that their motion would arrest after traversing a distance that is a few multiples of $\Delta_x$.
This is the reason for accumulation of large particles at a distance offset to the walls. If the time scale of shear-induced segregation with a shear rate $\dot{\gamma}$ is ${\text O}(1/\dot{\gamma}) \propto \sqrt{W/g}$, it is higher than the time scale for the free surface flow $\sqrt{(d_E/2)/g}$. The former implies that the accumulated large particles are advected at a faster rate with the downward flowing stream before they experience shear-induced segregation \citep{fan2011phase,neveu2022particle}, resulting in reverse segregation pattern as reported in figure~\ref{fig:noinsert}a. 

However, it should not be interpreted that the contribution from shear-induced segregation is negligible.  We state that rolling and bouncing-induced segregation can be the plausible mechanism in the present scenario that can dominate over shear.   The value of $\theta$ is an important flow-parameter here, and depends on the height of the feed stream and the width of the channel. For a large value of $\theta = 30^{\text o}$, the probability would be higher for both large and small particles  accumulating near the walls in the narrow channel due to continuous bouncing. However, in a wider channel, only large particles are expected to accumulate near the walls, as observed in earlier studies \citep{fan2012stratification,zhang2018size,isner2020axisymmetric}. The topic of free surface segregation  merits more attention to investigate the interplay among the angle of inclination of the free surface $\theta$ and the width of the channel $W/d_p$, and the velocities and fluxes of the particles (leaving from the edges). This forms a part of future work to understand free surface segregation  for the current flow configuration.

\subsection{With cylindrical inserts}
\label{sec:res_2}

\subsubsection{With one insert}
\label{sec:res_2_1}

\begin{figure}
\centering
\includegraphics[width=0.7\linewidth]{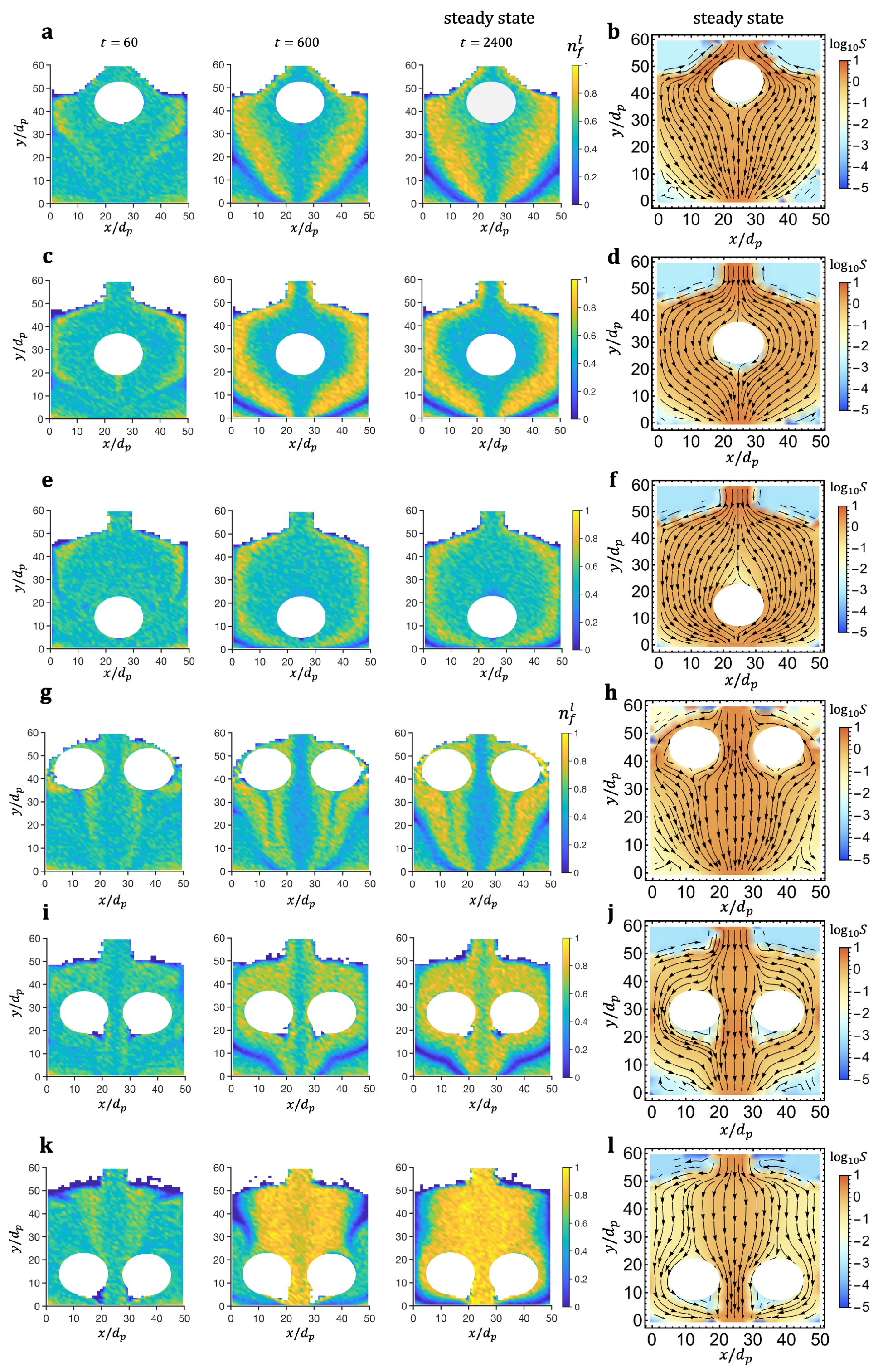}
\caption{With inserts.  The spatial distributions of large particles at different  times for one and two inserts at different depths are shown in (a), (c), (e), (g), (i) and (k); the background color represents the number fraction $n_f^l$ of large particles.   Streamlines of the flows at steady state are shown in (b), (d), (f), (h), (j) and (l); the background  color represents the value ${\rm{log}_{10}}S$, where $S = (u_x^2 + u_y^2)^{1/2}$ is the magnitude of the flow. The profiles in (a)--(f) are for one insert, and (g)--(l) for two inserts. In (a), (b), (g) and (h), $L/d_p = 45$;  in (c), (d), (i) and (j), $L/d_p = 30$; in (e), (f), (k) and (l), $L/d_p = 15$.}
\label{fig:inserts_seg}
\end{figure}

The degree of segregation reduces with the depth of the insert from the free surface (figures~\ref{fig:inserts_seg}a--f). Rolling and bouncing induced segregation is unavoidable due to the free surface at the top.  
When the insert is close to the free surface at $L/d_p = 45$, the segregation pattern and downstream flow profile appear similar to that without an insert (figures~\ref{fig:inserts_seg}a,b). It is noticeable that the angle of inclination of the free surface near the center of the channel is higher due to the insert. Therefore, large particles can reach the walls via continuous rolling or bouncing, resulting in a minor shift of the bands  towards the walls.  

\begin{figure}
\centering
\includegraphics[width=1\linewidth]{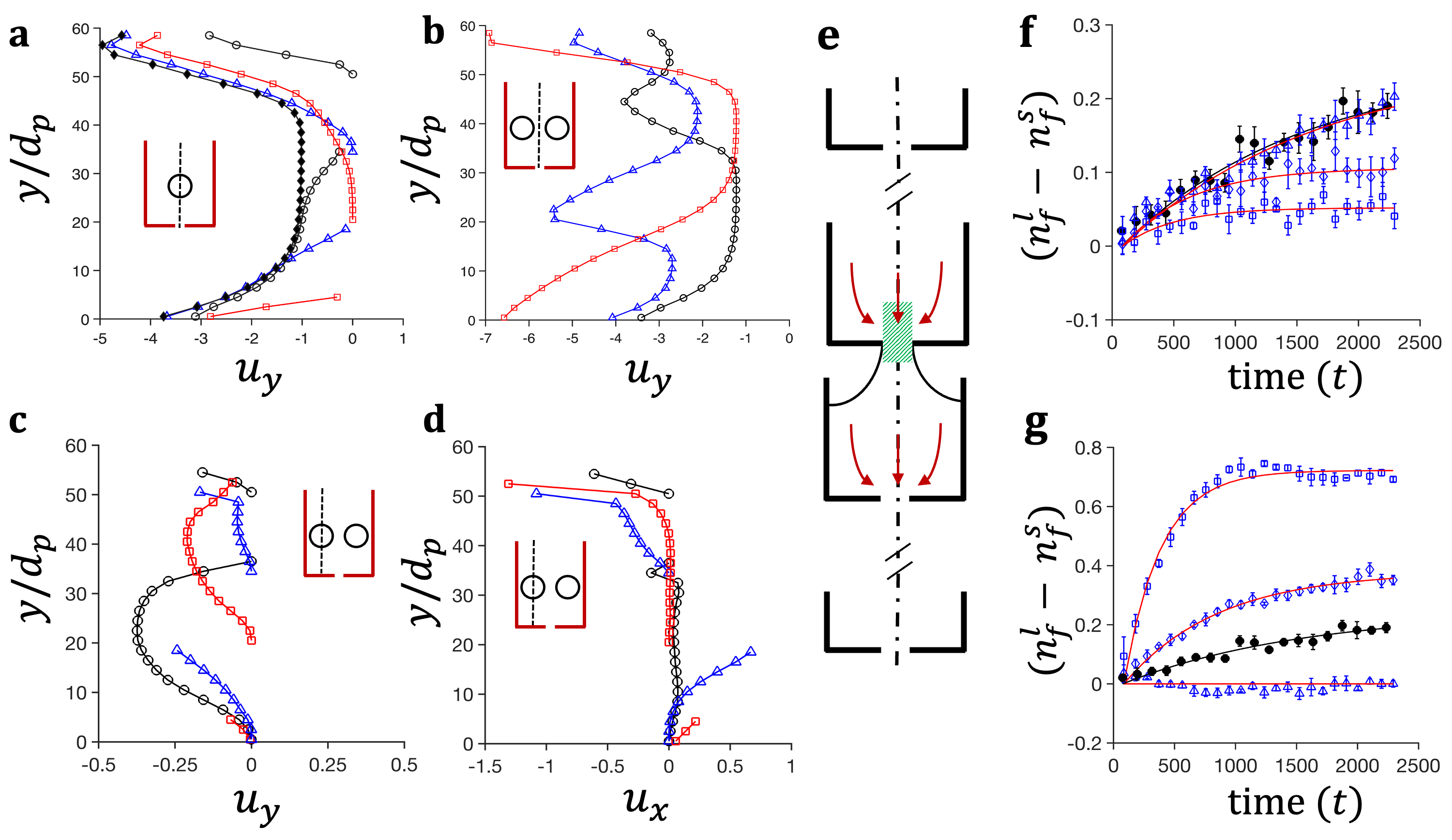}
\caption{Variation of flow velocity $u_y$ with $y/d_p$ at the mid-plane $x/d_p = 25$ for one insert (a) and two inserts (b). (c) and (d) With two inserts: variation of $u_y$ and  $u_x$ with $y/d_p$ at position $x/d_p = 12.5$.  In (a), the filled $\diamond$ symbol represents the data for the case--no insert. With one insert and two inserts at different heights in (a)--(d), black $\circ$, $L/d_p = 45$; blue $\triangle$, $L/d_p = 30$; red $\square$, $L/d_p = 15$. The data are shown for the planes marked as vertical lines in the insets. (e) Schematic of a multi-tray vertical tower. Variation of $(n^l_f - n^s_f)$ with time $t$ in the green shaded exit region shown in (e) for the cases of one insert (f) and two inserts (g). In (f) and (g), solid $\circ$, no insert; $\triangle$, $L/d_p = 45$; $\diamond$, $L/d_p = 30$; $\square$, $L/d_p = 15$. The error bars represent 95\% confidence limits of time-averaged values over a period $\Delta t = 125$. The red curves are the fits for the exponential function $y  =  b   (1 - e^{-at} )$.}
\label{fig:vel_midplane}
\end{figure}

When the insert is at an intermediate depth $L/d_p = 30$, bands of large particles shift towards the walls (figure~\ref{fig:inserts_seg}c). However, for $L/d_p = 15$, these bands almost disappear (figure~\ref{fig:inserts_seg}e). The streamlines  in figure~\ref{fig:inserts_seg}d indicate that the diversion in the flow from vertical to horizontal is stronger, shifting the segregated bands towards the walls. 
For $L/d_p = 15$, the flow diversion from vertical to horizontal is not strong enough to shift the bands (figure~\ref{fig:inserts_seg}f), unlike in $L/d_p = 30$.  In $L/d_p = 15$, the vertical velocity $u_y$ gradually decreases to zero above the insert at the mid-plane (figure~\ref{fig:vel_midplane}a). (The magnitude of the horizontal velocity  is negligible at this plane and is not shown.) This confirms the presence of a heap when the insert is close to the exit slot and no heap formation occurs in other cases. This heap formation is due to the pressure caused by the shock in the upstream region and friction which promote the material to pile above the insert. Such observations have been reported earlier \citep{gray2003shock,mathews2022numerical}. This flow modification due to the heap is significant, which leads to dissolution of the bands of large particles and stagnant zones (figures~\ref{fig:inserts_seg}e,f).  

The flow geometry and boundary conditions examined here equivalently represent the flow through a multi-tray vertical column (figure~\ref{fig:vel_midplane}e). The tray of interest is far from the trays at the top and bottom of the column. The exit stream from one tray is the inlet of the next tray. The difference in the number fractions between large and small particles, $(n^l_f - n^s_f)$, of the exit zone shown by the green shades in figure~\ref{fig:vel_midplane}e has been plotted over time. This quantity implies the extent of segregation; the values $(n^l_f - n^s_f) = $ 0 and 1 correspond to well-mixed and purely concentrated state with large particles, respectively. In figure~\ref{fig:vel_midplane}f, the profile for $L/d_p = 45$ is closer to the case with no insert. At other heights, the extent of segregation is reduced with a decrease in $L/d_p$. The trends of $(n^l_f - n^s_f)$ are shown to be well fitted by exponential functions, as explained by using a model in section~\ref{sec:res_2_3}. 

\begin{figure}
\centering
\includegraphics[width=1\linewidth]{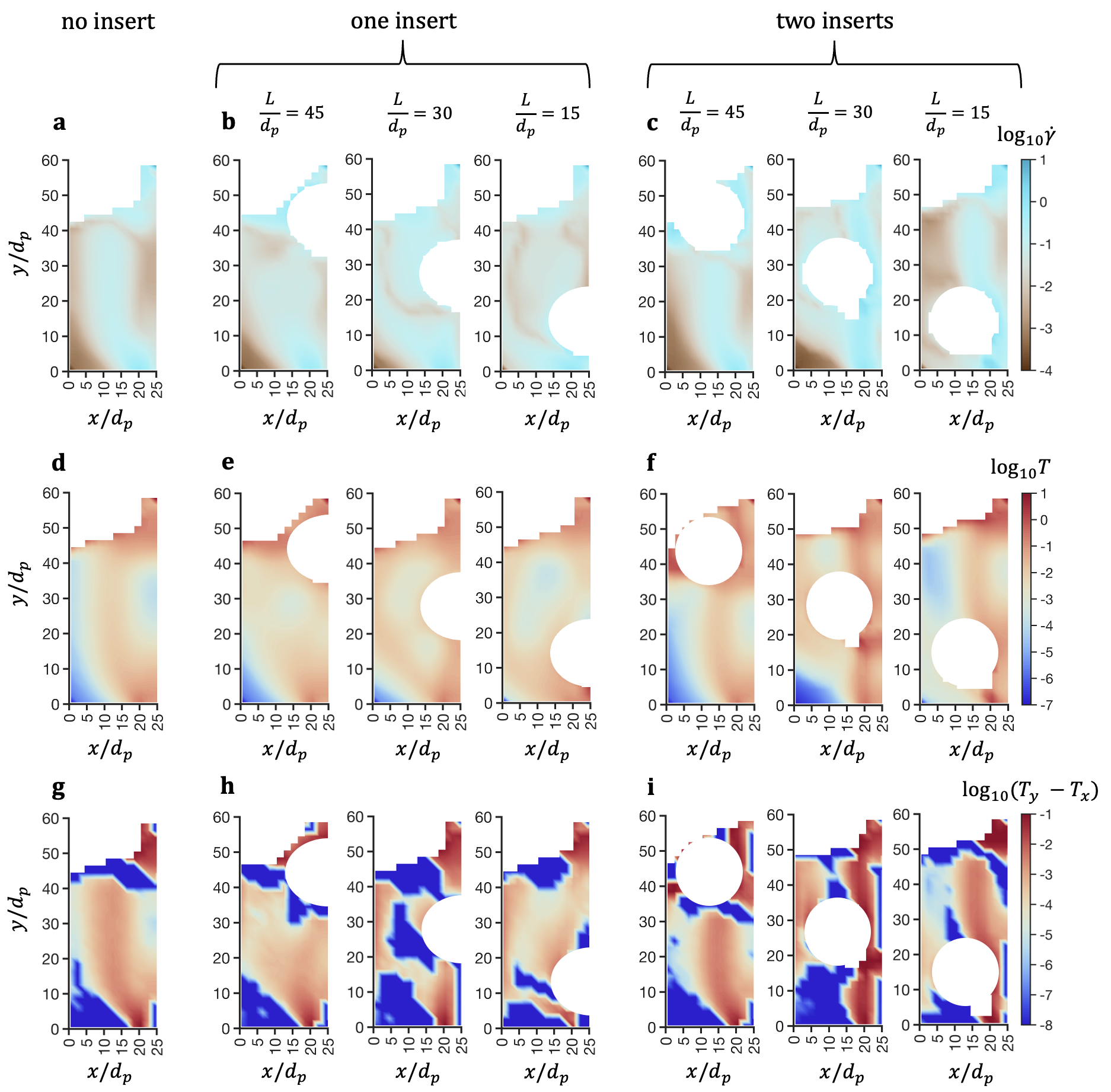}
\caption{The color maps of the shear rate $\dot{\gamma}$ ((a)--(c)), granular temperature $T$ ((d)--(f)), and difference in temperature $(T_y - T_x)$ ((g)--(i)) between $x-$ and $y-$ directions for different cases. The background color represents the $\rm{log_{10}(\star)}$ values of the above variables. No insert;  (a), (d), (g). With one insert for different values of $L/d_p$; (b), (e), (h). With two inserts for different values of $L/d_p$; (c), (f), (i).}
  \label{fig:temp_shear}
\end{figure}

The shear rate $\dot{\gamma}$ and granular temperature $T$  indicate that the gradients are strong in  the wall regions  when there is no insert and an insert close to the free surface i.e., $L/d_p = 45$ (figures~\ref{fig:temp_shear}a,b,d,e). The difference $(T_y-T_x)$ in the flow direction is also higher in regions where $\dot{\gamma}$ and $T$ are high (figures~\ref{fig:temp_shear}g,h). The quantity $(T_y-T_x)$ is an indirect measure of the degree of anisotropic diffusion, which promotes mixing in certain directions. The gradients in $\dot{\gamma}$ and $T$ weaken when the insert is inside the bed. The flow enhancement due to an insert inside a narrow channel homogenizes the fields around the insert. It is further strengthened when there is a heap above the insert in $L/d_p = 15$. We speculate that the modified flow pattern weakens the anisotropic diffusion field around the insert where $\dot{\gamma}$ and $T$ are finite, and  reverses the direction of the percolation flux at locations prone to segregation, resulting in enhanced mixing.  

\subsubsection{With two inserts}
\label{sec:res_2_2}

In figures~\ref{fig:inserts_seg}g,h and figures~\ref{fig:temp_shear}c,f,i, when the inserts are close to the free surface ($L/d_p = 45$), they tend to weaken the formation of strong bands of large particles. The stream that bifurcates above one of the inserts converges laterally and merges downstream, leaving almost no wake zone. Beneath the insert, where two streams rejoin, one stream  from the low shear region near the mid-plane brings higher momentum than the stream near the wall. Although $\dot{\gamma}$, $T$ and $(T_y - T_x)$ profiles in the downstream location are almost similar to those without insert (figures~\ref{fig:temp_shear}c,f,i), two counteracting streams with different momenta appear to inhibit strong band formation of large particles. This is in contrast with the cases with no insert or  single insert near the free surface. 

A striking observation is that the degree of reverse segregation is significantly enhanced compared to all other cases examined here, particularly when the inserts are close to the exit slot at $L/d_p = 15$ (figures~\ref{fig:inserts_seg}i--l). To the best of our knowledge, this finding has been reported for the first time. The flow is not symmetric around an insert. It is important to note that the speed of the material is significantly reduced near the wall regions (figures~\ref{fig:inserts_seg}j,l). The vertical velocity $u_y$ at the mid-plane is  higher than at the plane that passes through the center of the insert (figures~\ref{fig:vel_midplane}b,c). At height $L/d_p = 15$, heaps are forming above the inserts, which are absent for $L/d_p = 30$. This is further confirmed by the velocity profiles shown in figures~\ref{fig:vel_midplane}c,d. The profiles of $u_y$ and $u_x$ gradually decrease to zero as the flow approaches the inserts in $L/d_p = 15$, while only $u_y$ decreases to zero with non-zero $u_x$ above the inserts in $L/d_p = 30$. The occurrence of slow mobile zones near the walls is reflected in the profiles of $\dot{\gamma}$, $T$ and $(T_y - T_x)$ (figures~\ref{fig:temp_shear}c,f,i). The evolution of $(n^l_f - n^s_f)$ with time reflects the degree of enhanced segregation compared to that when there is no insert and the inserts are close to the free surface (figure~\ref{fig:vel_midplane}g) (the reason for exponential trends are discussed in section~\ref{sec:res_2_3}).  When the streamlines bifurcate above the inserts, the grains following the streamlines close to the wall lose their momentum due to increased frictional dissipation aided by narrow constrictions. The smallest gap for the material to flow in these regions is a few particle diameters that can even cause jamming. This effect is further enhanced due to the formation of heaps in the case $L/d_p = 15$. 

In these former cases, segregation due to the rolling and bouncing above the free surface is unpreventable. The small particles leaving the inlet stream reach the walls by bouncing. Their momentum dissipates near the wall regions, causing them to move slowly downward. The resistance to diffuse from the wall regions towards the central region increases as a result of these low-mobility regions, making it nearly impossible for small particles to rejoin the exit stream. As the flow is continuous due to the periodic boundary in the flow direction, small particles are continuously expelled from the inlet stream with negligible feedback in the exit stream. Thus, thin regions near the walls concentrated with small particles. Further the location of heaps in $L/d_p = 15$ appears to bifurcate the stream--one concentrated with small particles slowly moving near the walls and the other with large particles flowing relatively easily in the central zone. This is more likely the reason for enhanced segregation of large particles in the reverse direction towards the central region when inserts are close to the exit slot.

\subsubsection{Exponential scaling is explained by using a continuum model}
\label{sec:res_2_3}

The trends of $(n^l_f - n^s_f)$ are shown to be well fitted by exponential functions (solid red curves in figures~\ref{fig:vel_midplane}f,g). The reason for the exponential trends is the following. The time evolution of the number fractions of large ($n^l_f$) and small ($n^s_f$) particles are \citep{fan2014modelling,kumawat2025transient}
\begin{align}
    \frac{\partial n^l_f}{\partial t} \, + \,   \nabla \cdot (\mathbf{u} \, n^l_f)  \, + \, \nabla \cdot \mathbf{J}^l_{sg} \, + \, \nabla \cdot \mathbf{J}^l_d \, = \,  0 , \label{eq:balance_l}\\
    \frac{\partial n^s_f}{\partial t} \, + \,   \nabla \cdot (\mathbf{u} \, n^s_f)  \, + \, \nabla \cdot \mathbf{J}^s_{sg} \, + \, \nabla \cdot \mathbf{J}^s_d \, = \,  0.
\label{eq:balance_s}
\end{align}
Here $\mathbf{u}$ is the velocity of the mixture, and $\mathbf{J}^l_{sg}$ and $\mathbf{J}^l_d = D \nabla n^l_f$ are the segregation and diffusion fluxes, respectively, where $D$ is the diffusivity; superscripts $l$ and $s$ represent large and small particles, respectively. Consider the exit region. Subtracting (\ref{eq:balance_s}) from (\ref{eq:balance_l}), and neglecting the variations in the $x-$ direction, except the segregation flux which is assumed to be negligible in the $y-$ direction, 
\begin{equation}
    \frac{\partial}{\partial t}(n^l_f - n^s_f) \, = \, -\frac{\partial}{\partial y}\Big[u_y \,  (n^l_f - n^s_f) \Big] \, - \, \frac{\partial}{\partial x}(J^l_{sg,x} - J^s_{sg,x}) \, - \, \frac{\partial}{\partial y}(J^l_{d,y} - J^s_{d,y}). 
    \label{eq:bal_subtr}
\end{equation}
Neglecting the variation of number fractions in the flow ($y-$) direction, and therefore, omitting the term related to  diffusion $(J^l_{d,y} - J^s_{d,y} \, = \, D \, \frac{\partial}{\partial y}(n^l_f - n^s_f) = 0)$,
\begin{equation}
    \frac{\partial}{\partial t}(n^l_f - n^s_f) \, = \, -\, (n^l_f - n^s_f) \, \frac{\partial u_y}{\partial y} \, - \, \frac{\partial}{\partial x}(J^l_{sg,x} - J^s_{sg,x}). 
    \label{eq:bal_subtr_2}
\end{equation}
The above (\ref{eq:bal_subtr_2}) is equivalent to 
\begin{equation}
    \frac{\mbox{d}}{\mbox{d}t} (n^l_f - n^s_f) \, = \,  - \, m_1 \, (n^l_f - n^s_f) \, + \,  m_2,
    \label{eq:expon}
\end{equation}
if $m_1 = \frac{{\rm d} u_y}{
{\rm d} y}$ and $m_2 = -\frac{{\rm d}}{
{\rm d} x}(J^l_{sg,x} - J^s_{sg,x})$ are non-zero constants. Therefore, using
$(n^l_f - n^s_f) = 0$ at $t = 0$ as the initial condition, (\ref{eq:expon}) yields an exponential function
\begin{equation}
    (n^l_f - n^s_f) \, = \,  \frac{m_2}{m_1} \, (1 \, - \, e^{-m_1t}),
    \label{expon_fit}
\end{equation}
that fits the data well for $m_1 > 0$ and $m_2 > 0$ (red curves in figures~\ref{fig:vel_midplane}f,g). If $m_2 = 0$, ($n^l_f - n^s_f) = 0$. The profiles for $u_y$ in the exit region (figures~\ref{fig:vel_midplane}a,b) show that the slope is approximately constant  and positive. Therefore, to reduce (\ref{eq:bal_subtr_2}) to (\ref{eq:expon}), treating $m_1 > 0$ as a constant is a reasonable assumption. It is difficult to comment {\it a priori} on whether $m_2$ would be constant and positive. However, this choice seems to capture the trends satisfactorily.

\section{Discussion}
\label{sec:dis}

Mixing and segregation of  granular mixtures are energy intensive processes, for which gravity flow provides efficient solution. Since decades, a variety of  gravity flow setups have been engineered based on free-surface and shear-induced segregation mechanisms, enabling control over mixing and segregation through flow parameters such as overburden and flow rate of the material. In this work, we demonstrated that mixing and segregation can be controlled without modifying any flow parameters by introducing flow-modifying inserts and exploiting the confinement effects in a gravity flow setup. By changing number of inserts and their position, we showed significant enhancement in mixing and segregation. 

We investigated the flow of bi-disperse material through a narrow vertical channel with an exit slot using DEM simulations. We applied periodic boundary conditions in the flow direction to mimic the flow through a series of bins, and in the vorticity direction.  It has been observed that, without any insert, large particles accumulate between the central and wall regions, in contrast to the outcome in a purely shear-driven process \citep{fan2011phase}. The central region remains less affected (well-mixed), and the fraction of small particles is higher in the wall regions. Close examination revealed the influence of the free surface on segregation. But, in free surface segregation, it is more likely that large particles accumulate near walls \citep{samadani1999size}. We showed that rolling and bouncing induced segregation followed by downward advection is the dominant mechanism.  

Our phenomenological model predicted the existence of multiple regimes. In edge-rolling regime 1, a particle performs a continuous rolling motion, whereas the rolling motion ceases after traversing a distance in edge-rolling regime 2. In the rest of the regimes, a particle performs a projectile-like motion or bounces continuously until it dissipates its kinetic energy subsequently and falls under edge rolling regimes. It has been shown that a large fraction of large particles leaving the inlet stream fall under edge rolling regime 2, whereas small particles perform projectile-like motion or bounce continuously. This model successfully explained how rolling and bouncing on a low inclined free surface in a narrow column augment the accumulation of small particles in the wall regions and large particles away from the walls. The rolling and bouncing mechanism has been known for some time \citep{drahun1983mechanisms,fan2012stratification}. However, its strong impact on  segregation bands has been demonstrated for the first time in a gravity flow with a simple model. More work is required to examine parameters such as the width of the channel and width and height of the inlet stream, to identify the range when the rolling and bouncing mechanism is dominant. This is a separate topic in it-self and forms a part of future work.

With an insert inside the column, the flow pattern  changes significantly with an increase in its depth inside the bed. When the insert is close to the free surface, the downstream flow pattern is less affected. The former exhibits a segregation pattern and degree of segregation similar to that  without insert. When the insert is away from the free surface and the exit slot, the segregated bands of large particles are advected towards the wall, but a reduction of the degree of segregation in the exit stream. The shift in bands towards walls is due to sharp diversion in the flow. When the insert is close to the exit slot and far away from the free surface, the upstream pressure supports the material piling up above the insert. This is absent in the other cases due to lack of space required for a stable heap formation above the insert. The heap further modifies the flow substantially throughout the domain, and the degree of segregation was found to be further reduced. We hypothesize that the flow modification reverses the percolation flux, which mitigates band formation in segregation-prone regions. 

With two inserts, the degree of segregation is significantly enhanced when they are placed inside the column away from the free surface. When the inserts are close to the exit slot, the material piles up above the inserts, leaving the wall regions less mobile in the narrow channel. The material near the wall regions experiences a higher frictional resistance to flow than that in the central region, especially the small particles that accumulate near the walls via rolling and bouncing. The less mobile regions do not support them to migrate from wall regions to the central region and join the exit stream. The small particles are continuously expelled from the flowing stream, leaving the stream largely constituted by large particles. This leads to enhanced size segregation in the case of two inserts close to the exit slot. 

This work leads us to conclude that the cohesionless granular flow around an intruder or obstacle in a narrow channel provides enriching insights into local hydrodynamics and flow-structure interactions. There is a body of work on intruder dynamics. However, it is less often envisaged how such complex interactions can strongly influence material transport mechanisms under confined flow conditions.  These findings encourage to design and explore flow-structure interactions in confined dense granular flows to manipulate the material transport mechanisms. In complex flows in complex geometries, it is preferred to use the continuum approach to design and optimize the parameters. We did not rigorously attempt to use the former. While we believe that it will be fruitful to work with a continuum description in such scenarios to resolve complex flow patterns, the challenge may remain to choose or develop a well-posed constitutive theory and suitable boundary conditions for granular flows \citep{debnath2022comparison}.

\subsubsection*{Acknowledgements} The author thanks Prof. K. Kesava Rao and Prof. V. Kumaran for valuable discussions. The author acknowledges the support of PARAM PRAVEGA--IISc Bengaluru and PARAM SEVA--IIT Hyderabad under National Supercomputing Mission (NSM)
for providing computational support to generate the simulation data. B.D. acknowledges the Department of Science and Technology (DST), Ministry of Science and Technology, India
INSPIRE Faculty award DST/INSPIRE/04/2024/000725 for funding.

\bibliographystyle{jfm}
\bibliography{References}

\end{document}